\documentstyle[11pt]{article}
\makeatletter                                                           
\def\section{\@startsection {section}{1}{\z@}{-1.5ex plus -.5ex         
minus -.2ex}{1ex plus .2ex}{\large\bf}}                                 
\def\@thmcountersep{}                                                   
\def\ps@headings{                                                       
 \ifnum\thepage<10\def\@oddhead{\rm\hfill -- \thepage\ -- \hfill}       
 \else\def\@oddhead{\rm\hfill --\ \thepage\ -- \hfill}\fi               
 \def\@evenhead{\@oddhead}                                              
 \def\@oddfoot{}\def\@evenfoot{\@oddfoot} }      
\makeatother                                                            
\newfont{\bg}{cmr10 scaled\magstep5}
\newcommand{\bigzerol}{\smash{\hbox{\bg 0}}}
\newcommand{\bigzerou}{\smash{\lower1.7ex\hbox{\bg 0}}}
\newcommand{\re}{\par\hangindent=0.5cm\hangafter=1\noindent}

\newcommand{\etal}{{\rm et al.~}}

\newcommand{\alt}{\mbox{\raisebox{0.3ex}{$<$}\hspace{-1.1em}
                         \raisebox{-0.7ex}{$\sim$}} }
\newcommand{\agt}{\mbox{\raisebox{0.3ex}{$>$}\hspace{-1.1em}
                         \raisebox{-0.7ex}{$\sim$}} }

\newcommand{\dif}[2]{{\partial #1 \over \partial #2}}

\newcommand{\fracd}[2]{{\displaystyle{#1} \over \displaystyle{#2}}}

\newcommand{\ApJ}{ApJ}

\newcommand{\MN}{MNRAS}
\newcommand{\AAp}{A\&A}
\newcommand{\PASJ}{PASJ}
\begin{document}
\title{Collapse and Fragmentation of Magnetized Cylindrical Clouds\thanks{
Submitted to {\it \ApJ}}}
\author{Kohji T{\sc omisaka}\thanks{
        e-mail address: tomisaka@ed.niigata-u.ac.jp.}\\
	Faculty of Education, Niigata University,\\
        8050 Ikarashi-2, Niigata 950-21, Japan\\
	}
\date{Received: April 16, 1993; accepted:\hspace*{25mm}}

\maketitle\thispagestyle{empty}
\begin{abstract}
\noindent
Gravitational collapse of the cylindrical elongated cloud is studied
 by numerical magnetohydrodynamical simulations.
In the infinitely long cloud in hydrostatic configuration, small perturbations
 grow by the gravitational instability.
The most unstable mode indicated by a linear perturbation theory
 grows selectively even from a white noise.
The growth rate agrees with that calculated by the linear theory.
First, the density-enhanced region has an elongated shape, i.e., prolate
 spheroidal shape.
As the collapse proceeds, the high-density fragment begins to contract
 mainly along the symmetry axis.
Finally, a spherical core is formed in the non-magnetized cloud.
In contrast, an oblate spheroidal dense disk is formed in a cloud in which
 the magnetic pressure is nearly equal to the thermal one.
The radial size of the disk becomes proportional to the
 initial characteristic density scale-height in the $r$-direction.
As the collapse proceeds, a slowly contracting dense part is formed
 ($\alt$ 10\% in mass) inside of the fast contracting disk.
And this is separated from other part of the disk whose inflow
 velocity is accelerated as reaching the center of the core.
{}From arguments on the Jeans mass and the magnetic critical mass,
 it is concluded that the fragments formed in a cylindrical
 elongated cloud can not be supported
 against the self-gravity and it will eventually collapse.

\noindent
Subject Headings: Interstellar: Matter --- Interstellar: Magnetic Fields ---
Hydromagnetics --- Stars: Formation
\end{abstract}

\pagestyle{headings}
\section{Introduction}
The process of star formation begins in the interstellar clouds
 as a fragmentation of the clouds.
Massive stars are formed only in giant molecular clouds, while
 less massive stars are born also in less massive dark clouds
 (for a review, see Larson 1991).
This observational fact seems to suggest that
 the process of massive star formation and that of less massive stars
 are different.
This seems to be related to the difference between
 the collapse of supercritical clouds and subcritical clouds
 (Shu, Adams, \& Lizano 1987).
If a mass of the cloud is larger than a critical mass (for a magnetic cloud,
 $\sim$ magnetic flux/$G^{1/2}$),
 there is no equilibrium for the thermal pressure, the Lorentz force,
 and the centrifugal force to counter-balance the self-gravity.
The supercritical clouds begin dynamical collapse.
However, the subcritical magnetohydrostatic cloud is thought to evolve
 slowly only by the plasma drift (ambipolar diffusion; Mouschovias 1977,
 Nakano 1979) and/or the magnetic braking (Mouschovias 1979).
The structure of the cloud changes in a relatively long time scale
 of the plasma drift and the magnetic braking.
Finally, if the cloud becomes supercritical due to the decrease of magnetic
 flux at the center of the cloud or the decrease of angular momentum to support
 the cloud, it begins to dynamical collapse, too.
However, this is true only when the cloud in a static equilibrium is stable.
When the time scale of some dynamical instabilities is shorter than
 the evolutionary time scale, the cloud may begin a dynamical
 contraction or fragmentation
 before it reaches the condition for the supercritical cloud.
In a series papers, the fragmentation process in a subcritical,
 magnetohydrostatic cloud is studied by a non-linear magnetohydrodynamical
 simulation in two-dimension.
This paper is devoted to the filamentary cloud and its gravitational
 fragmentation.

The process of fragmentation and thus gravitational collapse of
 the high-density portion of the cloud are studied by a linear
 and non-linear analyses.
Linear analyses of the gravitational instability in an isothermal slab
 have been done many authors (see Spitzer 1978 for a review).
The qualitative answer is as follows: the slab is unstable for a perturbation
 whose wavelength is longer than a critical length and the most
 unstable perturbation has the wavelength of
 $\lambda_{\rm max}\simeq 20 c_s/(4\pi G \rho_c)^{1/2}$
 and the e-growing time scale of $\tau_{\rm max} \sim 2 (4\pi G
\rho_c)^{-1/2}$,
 where $\rho_c$ represents the density on the midplane of the disk
 (Elmegreen \& Elmegreen 1978).
The cylindrical cloud has a similar characteristic wavelength,
 $\lambda_{\rm max}\simeq  20 c_s/(4\pi G \rho_c)^{1/2}$
 and a longer growth time scale as
 $\tau_{\rm max}\sim 3 (4\pi G \rho_c)^{-1/2}$ (Nagasawa 1987).
It is concluded that
 if there is an inhomogeneity with an amplitude of $\Delta \rho / \rho
 \sim 5\%$, in $3 \tau_{\rm max} \alt 10 (4\pi G \rho_c)^{-1/2}$ the
 dense part of the cloud grows as a fragment with non-linear density contrast
 $\Delta \rho / \rho \sim 1$.
This time scale is comparable to that for the ambipolar diffusion
 (Nakano 1988).
Therefore, the dynamical instability in the subcritical
 static cloud should be studied.

The isothermal non-magnetic cylindrical cloud has a critical
 mass per unit length
 beyond which no equilibrium is achieved as
\begin{equation}
 \lambda_c= \int_0^\infty 2 \pi \rho r dr = 2 \frac{c_s^2}{G},
 \label{eqn:mcr-cyl}
\end{equation}
where $c_s$ represents the isothermal sound speed in the cloud.
The cylindrical cloud with a mass per unit length, $\lambda=\lambda_c$,
 has the infinite density contrast between the center and the surface.
As for the spherical isothermal cloud, the situation is different;
the critical cloud has a finite density contrast of
 $\rho_c/\rho_s=14$ (Bonner 1956; Ebert 1955).
The non-magnetic critical mass is equal to
\begin{equation}
  M_c= 1.18 \frac{c_s^4}{p_{\rm ext}^{1/2} G^{3/2}},
\label{eqn:mcr-sph}
\end{equation}
where $p_{\rm ext}$ represents the external pressure on the surface
 of the cloud.
Finally, in the slab geometry, no critical column density of
 the self-gravitating disk exists.
These differences come from the geometry or the dimension of the system.
Generally, the magnetic field and the centrifugal force in rotating clouds
 have an effect of increasing the critical mass.
{}From studying magnetohydrostatic equilibrium, the critical mass of the
 magnetized cloud is obtained as
\begin{equation}
  M_{c~{\rm mag}}
     \simeq 1.4 \left\{ 1 - \left[ \frac{0.17}{dm/d(\Phi_B/G^{1/2})}
                            \right]^2 \right\}^{-3/2}
                            \frac{c_s^4}{p_{\rm ext}^{1/2} G^{3/2}},
\label{eqn:mcr-mag}
\end{equation}
where $dm/d (\Phi_B/G^{1/2})$ means the mass-to-magnetic flux ratio
 at the center of the cloud (Tomisaka, Ikeuchi, \&Nakamura 1988).
Further the rotating cloud has a larger critical mass as
\begin{equation}
  M_{c~{\rm rot}}
     \simeq \left[ M_{c~{\rm mag}}^2 + \left(\frac{4.8c_s j}{G}\right)^2
            \right]^{1/2},
\label{eqn:mcr-rot}
\end{equation}
where $M_{c~{\rm mag}}$ and $j$ represent, respectively,
 the mass which can be supported without any rotation [Eq.(\ref{eqn:mcr-mag})]
 and a specific angular momentum of the cloud (Tomisaka, Ikeuchi, \& Nakamura
 1989).
Anyhow, the cloud with $\lambda > \lambda_c$ or $M > M_c$ eventually
 collapses, unless the excessive mass is eroded by any processes.

Study of the dynamical evolution of the magnetized cloud has been
 restricted for supercritical clouds (Scott \& Black 1980;
 Black \& Scott 1982; Phillips 1986a, b; Dorfi 1982, 1989).
The supercritical cloud contracts as a whole and forms a dense contracting
 core inevitably.
The authors reported that they did not observed any fragmentation
 in the process of contraction.
This seems to correspond to the fact that free-fall time scale
 is shorter than the growing time scale of the gravitational instability
 (see above).
The situation seems to be much different for subcritical clouds.

As for the non-magnetic cylindrical clouds,
Bastien (1983), and Bastien \etal (1991) have studied
 the contraction of initially uniform cylindrical cloud
 with finite length.
They achieved some conclusions on the fate of the above clouds:
 for example,
 for the elongated cloud with the ratio of length to diameter
 of the cylinder $\agt 2$, using the initial Jeans number, $J_0$,
 which is defined as the ratio of gravitational to the thermal
 energies, the evolution is determined.
 When $J_0 \alt J_{2-frag}$, two subcondensations are formed.
 For more large Jeans number as $J_{2-frag} \alt J_0 \alt
 J_{spindle}$, these two subcondensations collapses into one object.
 For the extreme case $J_0 \agt J_{spindle}$, the cloud contracts
 onto a line and forms a spindle.
However, their simulation has a restriction that
 the initial state is far from the hydrostatic equilibrium.
Excess free energy is liberated in the process of the relaxation
 from the initial state to the equilibrium,
 which may act an important role.
To understand the relatively slow evolution in the subcritical cloud,
 we should take a (magneto-)hydrostatic configuration
 as the initial state for the simulation.

%
%
\section{Model and Numerical Method}
\setcounter{equation}{0}
%
%
\subsection{Initial Condition}
As described in the preceding section,
 we assume that the initial state is in a hydrostatic equilibrium.
Using the cylindrical coordinate ($z$, $r$, $\phi$),
if the initial cylindrical isothermal cloud is homogeneous in the
$z-$direction,
 the magnetohydrostatic equilibrium configuration is obtained by
 the equation of hydrostatic balance and the Poisson equation as
\begin{equation}
 -\dif{\psi}{r}-\frac{c_s^2}{\rho}\dif{\rho}{r}
 -\frac{1}{8\pi\rho}\dif{B_z^2}{r}=0,\\
 \label{eqn:static}
\end{equation}
\begin{equation}
 \frac{1}{r}\dif{}{r}\left(r\dif{\psi}{r}\right)=4\pi G \rho,
 \label{eqn:poisson}
\end{equation}
where $\psi$, $\rho$, {\bf B}, $c_s$, and $G$ represent, respectively,
 the gravitational potential, density, magnetic field, isothermal
 sound speed, and the gravitational constant.
To derive equation(\ref{eqn:static}), it is assumed there is no helical
 magnetic field component ($B_\phi=0$).
The density distribution depends upon that of the magnetic field $B_z$.
In this paper, we restrict ourselves to two cases:
 (i) uniform magnetic field $B_z=$constant, and
 (ii) the ratio of the magnetic pressure to the thermal one is constant.
In the first model, the density distribution becomes identical to that
 of nonmagnetic isothermal cylinder as
\begin{equation}
 \rho(r)=\frac{\rho_c}
        {\left(1+\fracd{r^2}{8}\fracd{4\pi G \rho_c}{c_s^2}\right)^2},
 \label{eqn:nomag}
\end{equation}
where $\rho_c$ means the density at the center of the cylinder ($r=0$).
As easily seen, the density reaches zero only at the infinity ($r=\infty$).
Thus the isothermal cloud should be bounded by the external pressure,
 $p_{\rm ext}$.
The cloud has a boundary where $\rho(r_s)$ is equal to
 $p_{\rm ext}/c_s^2$.
The cloud radius is expressed as
\begin{equation}
 r_s=2^{3/2}\left[\left(\frac{\rho_c}{\rho_s}\right)^{1/2}-1\right]^{1/2}
     \frac{c_s}{(4\pi G \rho_c)^{1/2}}.
 \label{eqn:r_s}
\end{equation}
where $\rho_s$ is the density on the cloud surface and equals to
 $p_{\rm ext}/c_s^2$.
In the present paper, we use the normalization as $c_s=4\pi G=p_{\rm ext}=1$.
Thus the unit of the distance is chosen as $H=c_s/(4\pi G \rho_s)^{1/2}$.
The normalized density distribution, $f(r)\equiv \rho(r)/\rho_s$,
 and the radius of the surface, $\xi_s\equiv r_s/H$, are expressed as
\begin{equation}
 f(\xi)=\frac{F}{\left[1+\fracd{F}{8}\xi^2\right]^2},
 \label{eqn:f}
\end{equation}
\begin{equation}
 \xi_s=\left(\frac{2^3}{F}\right)^{1/2}
       \left[F^{1/2}-1\right]^{1/2},
 \label{eqn:xi}
\end{equation}
where $\xi\equiv r/H$ represents the normalized distance and
 $F=\rho_c/\rho_s$ denotes the density contrast between the center
 and the surface.

When the magnetic field plays a role in supporting the cloud,
 the density distribution is different from equation(\ref{eqn:nomag}).
Assuming that the magnetic pressure is proportional to the thermal pressure,
 i.e., $B_z^2/8\pi \propto c_s^2 \rho$, the distribution of the density
 becomes as
\begin{equation}
 f(\xi)=\frac{F}{\left[1+\fracd{F}{8}\fracd{\xi^2}{1+\alpha/2}\right]^2},
 \label{eqn:fB}
\end{equation}
where the parameter $\alpha$ is defined as
\begin{equation}
 \alpha\equiv \frac{B_{z}^2/4\pi}{c_s^2 \rho}\equiv \frac{2}{\beta},
 \label{eqn:alpha}
\end{equation}
in terms of the plasma $\beta$.
This shows that the cloud becomes thick with increasing $\alpha$,
 and the density scale-height changes in proportion to $(1+\alpha/2)^{1/2}$.
Typical density distributions are shown in Figure 1.

We assumed a hypothetical situation that the cloud is confined in a
 low-density ambient medium which has no importance as the source of gravity
 but has a finite pressure $p_{\rm ext}$.
Thus, the initial solution of $\rho(r)$ has a boundary where
 $\rho(r)=p_{\rm ext}/c_s^2$.
Beyond this cloud boundary, a tenuous and thus hot medium is assumed to extend.
However, as seen in the next subsection, since we assume the isothermal
 equation of state, it is difficult to calculate two kinds of gases,
 cold cloud component and warm intercloud component, separately.
Therefore, we adopt a one-fluid approximation, where the gas has an identical
 temperature; but the ambient gas is defined as the portion with
 $\rho < p_{\rm ext}/c_s^2\equiv \rho_s$, and this gas has no effect of the
 gravitational field made by the cloud.
This gives a virtual distribution of ambient gas as
\begin{equation}
  f(\xi)=\left( \frac{\xi}{\xi_s} \right)^{-\frac{4(F^{1/2}-1)}{F^{1/2}}},
\label{eqn:ambrho}
\end{equation}
to counter-balance the gravity by the cloud as
\begin{equation}
  g(\xi)=\dif{\ln \rho}{\xi}=-\frac{4}{\xi}\frac{F^{1/2}-1}{F^{1/2}}.
\label{eqn:ambg}
\end{equation}
%
%
\subsection{Basic Equations}
The cylindrical symmetry is assumed: $\partial/\partial\phi=0$.
For the dense gas found in interstellar clouds with $\simeq 10$K,
 the equation of state is well approximated with the isothermal one.
We assume here that the gas obeys the isothermal equation of state.
Thus the basic equations become the equations of magnetohydrodynamics
 for isothermal gases as
\begin{equation}
 \dif{\rho}{t}+\dif{}{z}(\rho v_z)+\frac{1}{r}\dif{}{r}(r\rho v_r)=0,
 \label{eqn:continuity}
\end{equation}
\begin{equation}
 \dif{\rho v_z}{t}+\dif{}{z}( \rho v_z v_z) +
  \frac{1}{r}\dif{}{r}(r\rho v_z v_r)= - c_s^2 \dif{\rho}{z}
     -\rho\dif{\psi}{z}
     -\frac{1}{4\pi}\left(\dif{B_r}{z}-\dif{B_z}{r}\right)B_r,
 \label{eqn:vz}
\end{equation}
\begin{equation}
 \dif{\rho v_r}{t}+\dif{}{z}( \rho v_r v_z) +
  \frac{1}{r}\dif{}{r}(r\rho v_r v_r)= - c_s^2 \dif{\rho}{r}
     -\rho\dif{\psi}{r}
     +\frac{1}{4\pi}\left(\dif{B_r}{z}-\dif{B_z}{r}\right)B_z,
 \label{eqn:vr}
\end{equation}
\begin{equation}
  \dif{B_z}{t}=\frac{1}{r}\dif{}{r}[r(v_zB_r-v_rB_z)],
  \label{eqn:Bz}
\end{equation}
\begin{equation}
  \dif{B_r}{t}=-\dif{}{z}(v_zB_r-v_rB_z),
  \label{eqn:Br}
\end{equation}
\begin{equation}
  \frac{\partial^2 \psi}{\partial z^2}
 + \frac{1}{r}\dif{}{r}\left(r\dif{\psi}{r}\right) =
   4 \pi G \rho,
  \label{eqn:poisson2}
\end{equation}
 where the variables have their ordinary meanings.

To include only the gas density in the cloud as the source of
 the self-gravity, the right-hand side of the equation
(\ref{eqn:poisson2}) is calculated as follows:
\begin{equation}
 \rho(z,r)= \left\{ \begin{array}{ll}\rho, &\mbox{if $\rho > \rho_s$,} \\
                                      0,   &\mbox{if $\rho < \rho_s$.}
                    \end{array}\right.
\label{qn:varrho}
\end{equation}
As long as the cloud keeps its hydrostatic configuration,
 the solution of the above equations coincides with
 the initial configuration, equations (\ref{eqn:f})
 and (\ref{eqn:ambrho}).
However, if the ambient matter accretes onto the cloud by the gravity or
 the cloud material evaporates by the pressure force,
 the above procedure may give wrong results.
We monitored the mass of the cloud in numerical runs,
 and check the validity of the ambient matter condition.
We confirmed that the cloud mass only changed less than $\sim 3$\%.

Nondimensional variables are used as follows:
 the density, $\rho'\equiv \rho/\rho_s$,
 the velocity, ${\bf v}'\equiv {\bf v}/c_s$,
 the gravitational potential, $\psi'\equiv \psi/c_s^2$,
 the magnetic fields, ${\bf B}'\equiv {\bf B}/(4\pi c_s^2 \rho_s)^{1/2}$,
 the time, $t'\equiv t (4\pi G \rho_s)^{1/2}$,
 and the linear size, ${\bf r}'\equiv {\bf r}/H$.
Using this normalization,
 equations (\ref{eqn:vz}), (\ref{eqn:vr}),
 and (\ref{eqn:poisson}) become
\begin{equation}
 \dif{\rho' v'_z}{t'}+\dif{}{z'}( \rho' v'_z v'_z) +
  \frac{1}{r'}\dif{}{r'}(r'\rho v'_z v'_r)= - \dif{\rho'}{z'}
     -\rho'\dif{\psi'}{z'}
     -\left(\dif{B'_r}{z}'-\dif{B'_z}{r'}\right)B'_r,
 \label{eqn:vz-n}
\end{equation}
\begin{equation}
 \dif{\rho' v'_r}{t'}+\dif{}{z'}( \rho' v'_r v'_z) +
  \frac{1}{r'}\dif{}{r'}(r'\rho' v'_r v'_r)= - \dif{\rho'}{r'}
     -\rho'\dif{\psi'}{r'}
     +\left(\dif{B'_r}{z'}-\dif{B'_z}{r'}\right)B'_z,
 \label{eqn:vr-n}
\end{equation}
\begin{equation}
  \frac{\partial^2 \psi'}{\partial z'^2}
 + \frac{1}{r'}\dif{}{r'}\left(r'\dif{\psi'}{r'}\right) =
   \rho'.
  \label{eqn:poisson2-n}
\end{equation}
Other equations are identical even when these normalized variables are used.
Hereafter, we will use these nondimensional variables and
 abbreviate the prime unless otherwise mentioned.
%
%
\subsection{Numerical Method}

The basic equations are solved by the finite difference method.
The mesh spacing ($\Delta z$, $\Delta r$) is constant spatially
 except for models CB and CB2 in table 1.
Unequal spacing is used for model CB and CB2 to see the fragment more closely.
Cell numbers in one dimension are chosen $N=200-800$ (table 1).
The ``Monotonic Scheme'' (van Leer 1977; Norman \& Winkler 1986) is adopted
 to solve the hydrodynamical equations,
 i.e., equations (\ref{eqn:continuity})--(\ref{eqn:vr}),
 and the ``Constrained Transport'' method (Evans \& Hawley 1988)
 is adopted to solve the induction equations of the magnetic field,
 i.e., equations (\ref{eqn:Bz}) and (\ref{eqn:Br}).
Although the equation of state is different,
 this scheme is essentially the same as that used by Tomisaka (1992),
For numerical stability at the shock front, an artificial viscosity is
 included in equations (\ref{eqn:vz}) and (\ref{eqn:vr}).
A tensor artificial viscosity term is used as shown in the Appendix
 of Tomisaka(1992).
The program for adiabatic gas was checked by several test problems:
 Sedov solution, spherical
 stellar wind bubble solution (Weaver \etal 1977),
 one-dimensional magnetic Rieman problem (Brio \& Wu 1988).
Further, the contraction of a spherical isothermal cloud is compared
 with solutions of Larson (1969) and Penston (1969).
The radial density distribution is well fitted by $\propto r^{-2}$
 and derived velocity profiles agree with their results by one-dimensional
 calculation.

To solve the Poisson equation is reduced to find the solution of
 simultaneous linear equations with a dimension of $N^2$.
The linear equations are solved by the ``Conjugate Gradient Squared method''
 (CGS method, Dongarra \etal 1991)
 preconditioned by the modified incomplete LU decomposition (MILUCGS)
 (Meijerink \& van der Vorst 1977; Gustafsson 1978).
A brief description of this method and its performance are described
 in Appendix.

On the upper ($z=l_z$) and lower ($z=0$) boundary,
 the cyclic boundary condition is applied.
On the outer boundary the fixed boundary condition is applied.
%
%
\section{Result}
\setcounter{equation}{0}
%
%
\subsection{A Typical Evolution}

First, the numerical result without any initial perturbation is seen.
This corresponds to the model with $\alpha=1$ and
 $F\equiv c_s^2 \rho_c/p_{\rm ext}=\rho_c/\rho_s=100$ (model A).
The size of the calculated region is taken as $1.93 H$,
 which coincides with the wavelength which has the maximum
 growth rate in the gravitational instability
 (Nakamura, Hanawa, \& Nakano 1993).
Figure 2 shows the evolutions of density (left panel) and magnetic fields
 (right panel).
In $t=3.6$, no prominent fragmentation appears.
In the stage shown in Figure 2c, a high-density region elongating
 in the $z$-direction appears.
The shape of this high-density region is prolate, i.e.,
 if it fits to an ellipsoid the major axis coincides with
 the symmetric axis (the $z$-axis).
This agrees with the result of linear analyses (Nakamura \etal 1993;
 Nagasawa 1987);
The wavelength of the most unstable perturbation was predicted
 from the linear analysis as $\simeq 2\times H$.
And the global shape of the high-density region in Figure 2c coincides with
 an expected shape from the eigen function of the linear analysis.
{}From this figure, it is confirmed that
 the perturbation with a shorter wavelength
 than that of the most unstable mode does not appear at all, before
 the most unstable mode grows into the non-linear region.
However, a perturbation with a longer wavelength than the numerical
 box can not be properly calculated.
We have calculated in model A2 (not shown) the same problem with a
 larger spacing and thus with a numerical box twice as large as
 model A.
In model A2, two high-density regions appear and each fragment shows
 the identical evolution as model A, which confirms the most unstable
 mode expected from the linear analysis grows first.

After the stage in Figure 2c, collapse along the symmetry axis proceeds
 as well as the radial direction.
The final structure is a disk contracting towards the center seen
in Figure 2d.
Since the gas easily falls down along the magnetic field lines,
 this disk is perpendicular to the symmetric axis of the cylinder.
The magnetic fields are squeezed by the effect of radial contraction
 near the disk (right panel).
Since this configuration is unstable against the Parker instability (1979
 and references therein),
 the gas is accumulated flowing along the magnetic fields.
Contracting speed in the $z$-direction is $\simeq$ twice as fast as the
 radially contracting speed.
In the central region, a high-density core is formed, on to which
 accretion of extra mass continues.
Finally, near the high-density core the infalling speed exceeds the
 isothermal sound speed at this stage.
It is concluded that a typical product of the self-gravitational instability
 in the cylindrical
 cloud is oblate spheroidal disks separated by $\lambda_{\rm max}$.

The phase of the position where the disk is formed ($z_c$) is
 not determined a priori.
This is related to the question that where the irregularity
 comes from even if the initial state is uniform along the $z$-axis.
After several numerical tests, it is concluded that
 the main contribution of irregularity is coming from the Poisson
 solver;
Even if the density distribution, $\rho$, is uniform in the $z$-direction,
 the numerical values of the potential, $\psi$, is not completely
 uniform.
This is due to the CGS scheme.
Although the typical relative error in the potential is small,
 $\Delta \psi / \psi \sim 10^{-7}$,
 this irregularity grows by the effect of gravitational instability and
 finally it forms fragmentations.

To confirm the idea, another model B is calculated,
 whose parameters are identical
 with model A except for the existence of initial irregularity.
In model B, a perturbation with relative contrast $\delta_{\rm ini} \equiv
 \Delta \rho / \rho = 10^{-2}$ is added as
\begin{equation}
  \rho(z,r)=\rho_0(r) \left(1 - \delta_{\rm ini} \cos{2\pi z\over l_z}\right),
\end{equation}
where $\rho_0$ and $l_z$ represent, respectively,
 the initial density determined by $f(\xi)$ of equation(\ref{eqn:fB})
 and the $z$-length of numerical box.
The wavelength of this perturbation is again taken identical with that of
 the most unstable mode in the linear analysis.
The structure at $t=1.26$ is shown in Figure 3.
This shows that the final structure of this model is identical with
 that of the model A (Fig.2d), although the time necessary for the
 fragment to grow is shortened much (from $t=5.6$ to $t=1.26$).
It is clear that
 this quick evolution is due to a finite amplitude of the perturbation added
 to the initial state,
 compared with the intrinsic irregularity from
 the Poisson solver in model A.
However, the evolution itself is very similar with each other.

To confirm this, the time evolution of the amplitude of the
 density irregularity is plotted for both models A and B in
 Figure 4.
Using the average $<\rho(r=0)>$ and the maximum $\rho(r=0)|_{\rm max}$
 density on the $z$-axis, the relative density contrast
\begin{equation}
  \delta_{\rm max} \equiv \frac{\rho(r=0)|_{\rm max}}{<\rho(r=0)>}-1,
\label{eqn:delta}
\end{equation}
is plotted against the time passed after the calculation begins.
This shows that in model A the perturbation grows
 in accordance with $\propto \exp{[\omega t]}$ for $1 \alt t \alt 5$,
 from $10^{-8}$ to 0.1.
Model B shows the similar growth but the region in which it shows the
 exponential growth is much restricted.
However, if the origin of time for model B is moved,
 two curves for models A and B coincides.
This indicates that
 the evolution is similar whether the most unstable mode is overlaid
 on to the initial density distribution (model B) or not (model A).
This confirms the expectation that the most unstable mode grows
 selectively even from the irregularity of white noise.
Thus, we add the perturbation of the expected most unstable mode to the
 initial density distribution as model B.
%
%
\subsection{Non-Magnetic Cylinder}

To see the effect of the magnetic field,
 we calculated the non-magnetic isothermal cylinder as model C.
{}From the linear analysis, the wavelength of the most unstable mode
 is equal to $\simeq 2.21H$ (Nakamura \etal 1993).
Since the magnetic field has an effect of supporting the cloud,
 the radial size of the cloud becomes more compact than
 that of models A and B.

In Figure 5, the evolution of model C is shown.
In $t=1.2$, the fragment, whose density contrast $\delta$ reaches
 $\sim 3$, has entered in non-linear region (Fig.5a).
However, comparing with Figure 3 (these two models are assumed to have
 irregularities with the same initial amplitude), it is shown that
 this model evolves slower than model B.
This difference is understood as follows:
 (1) since the magnetic fields of model B
 has a configuration unstable for the Parker instability,
 the system becomes more unstable including the magnetic fields,
 or (2) the mass per unit length of the cylinder and thus
 the effect of the self-gravity becomes
 larger when the magnetic fields are supporting the cloud.

Figure 5b shows the structure at the age of $t=1.4$.
In this phase, the maximum density at the center of the
   core reaches $10^5$.
The envelope with low-density $\simeq 10^{1.5}$
 seems to have the shape of prolate spheroid as well as the magnetic models.
In contrast, the high-density central region appears to be
 almost spherical symmetric.
The difference between structures appearing in the magnetic cloud
 and the non-magnetic one is apparent.
Since there is no lateral restoring force by the magnetic fields,
 the high density fragment shows spherical shape, while in the
 magnetic cloud the fragment is formed from the matter flowing along
 the magnetic fields as a disk perpendicular to the
 magnetic fields.
%
%
\subsection{Central Core}

To see the structure of the central core,
 the density profile is plotted against
 $\log |z-z_c|$ and $\log r$ in Figure 6 for models B and C,
 where $z_c$ represents the position of the core center.

There exist two characteristic power law solutions.
One is a singular isothermal sphere solution (Chandrasekhar 1939):
\begin{equation}
 \rho_{\rm sing} = \frac{c_s^2}{2\pi G r^2},
 \label{eqn:singularS}
\end{equation}
or in a non-dimensional form:
\begin{equation}
 \frac{\rho_{\rm sing}}{\rho_s} = \fracd{2}{(r/H)^2}.
 \label{eqn:singularS'}
\end{equation}
The other is an asymptotic solution of equation(\ref{eqn:f})
 for $\xi \gg 1$ or $r \gg H$:
\begin{equation}
 \rho_{\rm out} = \fracd{\rho_c}{\fracd{r^4}{8^2}
                       \fracd{(4\pi G \rho_c)^2}{c_s^2}},
 \label{eqn:singularC}
\end{equation}
or in a non-dimensional form:
\begin{equation}
 \frac{\rho_{\rm out}}{\rho_s}=\frac{1}{F}\left(\frac{8}{\xi^2}\right)^2.
 \label{eqn:singularC'}
\end{equation}
As seen in Figure 6,
 the distribution in the $r$-direction is well fitted by
 equation (\ref{eqn:singularC'}) for the outer envelope.
As collapse proceeds, the radial density distribution
 in the inner part of the cloud becomes similar to
 that of the singular isothermal sphere
 [equation (\ref{eqn:singularS})].

As for the $z$-distribution,
 in the {\it non-magnetic cloud} (model C), the density distribution
 is very similar to that in the $r$-direction.
However, in the {\it magnetic cloud} (model B), the $r$- and
 $z$-distributions are much different.
This difference comes from the effect of magnetic fields,
 that is, in models A and B a disk is finally formed,
 although in model C an almost spherical core is formed.
Since the gravities by a disk and a sphere are quit different,
 the resultant density distributions are different.

When the spherical cloud has the density distribution like equation
 (\ref{eqn:singularS}), the total mass inside of the sphere $R$
 is written as
\begin{eqnarray}
 M(R) &=& \int_0^R \rho_{sing} 4 \pi r^2 dr = \frac{2c_s^2}{G} R,
                                                       \nonumber \\
      &=& \left(\frac{2}{\pi}\right)^{1/2}
             \frac{c_s^4}{[\rho_{sing}(R)c_s^2]^{1/2}G^{3/2}},
\end{eqnarray}
using the density $\rho(R)$ at $r=R$.
This represents the Jeans mass for the spherical cloud with
 infinite center-to-surface density contrast, $M_J(R)$.
Thus, when the density increases more rapid than equation
 (\ref{eqn:singularS}) toward the center,
 the actual mass is larger than this Jeans mass and
 contraction never stops.
In contrast, when the density distribution is flat,
 the actual mass $M(r)$ never surpasses $M_J(r)$
 in any region of $r < R$.
Considering this,
 since the radial density distribution $\rho \agt \rho_{sing}$,
 the spherical core formed in the center of the non-magnetic cloud
 seems to continue to contract.

{}From the studies on magnetohydrostatic equilibrium
(Tomisaka \etal 1988),
 the mass-to-flux ratio at the {\em center} of the cloud,
 $M/(\Phi/G^{1/2})$,
 has a crucial role to divide the clouds into super- and subcritical
 clouds;
that is, considering a cloud whose mass is larger than the non-magnetic
 critical mass ($\agt M_{\rm cr}$ of eq.[\ref{eqn:mcr-sph}]),
when the mass-to-flux ratio is larger than $\simeq 1/(2 \pi)$ the cloud
 has no equilibrium solutions;
while the cloud has at least one hydrostatic solution,
 when the ratio is smaller than this value.
Since the mass-to-flux ratio remains constant along the contraction,
 the mass-to-flux ratio {\it in the fragment} can be written using quantities
 in the initial stage as
\begin{eqnarray}
  \frac{M}{\Phi / G^{1/2}} &=& \frac{\rho_c l_z G^{1/2}}{B_c}
      =  \frac{\rho_c^{1/2} l_z G^{1/2}}{(4\pi\alpha)^{1/2}c_s},
\end{eqnarray}
where $l_z$ represents the $z$-length of the numerical box and
 is taken equal to the wavelength of the most unstable perturbation,
 $\simeq 20 c_s/(4\pi G \rho_c)^{1/2}$.
This ratio is approximately equal to
\begin{equation}
  \frac{M}{\Phi / G^{1/2}}\simeq \frac{1.59}{\alpha^{1/2}},
  \label{eqn:M/Phi}
\end{equation}
Thus, for the cloud with the mass appreciably larger than that of equation
 (\ref{eqn:mcr-sph}), only when the magnetic field is extremely strong as
 $\alpha \agt 100$, the magnetized fragment may be subcritical.
The contraction of the core can not be stopped by
 a magnetic field, which is assumed ordinarily, $\alpha\simeq 1$.
This is valid until the wavelength of the most unstable perturbation
 is much longer than $20 c_s/(4\pi G \rho_c)^{1/2}$ in a cloud
 with low density contrast $F$.

In conclusion, the cores formed in the fragment of the cylindrical
 cloud seems to continue to contract,
 until other effects not considered here may work in supporting the fragment,
 such as, the equation of state
 of the gas changes from the isothermal to more hard one.
%
%
\subsection{Diffuse Cloud}

Here, we compare the cloud with high central density and
 that with low density contrast calculated in models D.
In Figure 7a, the evolution of model DB is plotted.
Both models B and DB have the same parameters except for the
 center-to-surface density ratio, $F$.
Difference between models B and DB are apparent:
 although a disk is formed for both models by the effect of the magnetic field,
 the disk perpendicular to the magnetic fields of model DB is much larger
 than that of model B.
The disk has a radial size of 0.16 (model B) and 0.71 (model DB)
 (the size is measured as a radius of the largest contour line which has
 an oblate shape).

The radial extent of the finally formed disk seems to
 correspond to the
 size of the initial central cylindrical core, i.e.,
 in equation (\ref{eqn:nomag}) the initial density is almost uniform
 irrespective of the radial distance for
 $r \alt r_c = 2^{1.5}(1+\alpha)^{1/2} c_s/(4\pi G \rho_c)^{1/2}$.
Since this gives
\begin{equation}
 r_c \simeq 0.4 \left(\frac{F}{100}\right)^{-1/2}
                \left(\frac{1+\alpha}{2}\right)^{1/2}
                \frac{c_s}{(4\pi G \rho_s)^{1/2}},
\end{equation}
 the above result shows that a disk threaded by magnetic fields
 whose radial extent is equal to
 $\sim$ 40 -- 55 \% of the initial core size is formed.

If this is true,
 does the cloud with a large $\alpha$ form a large disk?
Yes.
In model DD with $\alpha=4$ (Fig.~7b),
 the disk with a radial size of $\sim 1.2$ is formed,
 while $r_c \simeq 2 (F/10)^{1/2} [(1+\alpha)/5]^{1/2}$.
Thus, it can be concluded that in the magnetized cylindrical
 cloud a disk is formed perpendicularly threaded by the magnetic
 fields and its radial extent is related to $r_c$.

The next characteristic point of the evolution of diffuse clouds
 is a low growth rate of fragmentation.
A typical time scale for the perturbation to evolve
 into the nonlinear stage is $t\sim 4.72$ for $\alpha=0$ (model DC),
 $t\sim 4.12$ for $\alpha=1$ (model DB),
 and $t \sim 3.31$ for $\alpha=4$ (model DD).
This is much longer than that of models B and C with $F=100$.
This apparently comes from the fact that the growth time-scale
 is proportional to the free-fall time for the central density.
Finally, comparing the non-magnetic clouds of models C and DC,
 Figures 4 (model C) and 7c (model DC) look very similar.
These figures are scaled in proportional to $l_z$,
 and $l_z$ is chosen as $\lambda_{\rm max}\simeq 20 c_s/(4\pi G\rho_c)^{1/2}$.
This means that the actual size of the spherical core of model DC is just
 $10^{1/2}$ times larger than that of model C.
If we scale the size by $c_s/(4\pi G\rho_c)^{1/2}$ instead of
 $H=c_s/(4\pi G\rho_s)^{1/2}$,
 the structures of these models becomes almost identical.
Therefore, the structure of the fragment is
 mainly determined by the initial highest density
 at the center of the cylinder cloud for the non-magnetic cloud.

\subsection{Further Evolution}

To see the structure of the contacting central core in more detail,
 we have to execute larger calculation with finer resolution
 using a plenty of meshes.
However, this approach will meet an inevitable limit of available memories
 on any computers.
Therefore, we avoid this difficulty by varying mesh sizes,
 as a finer mesh is used around the expected position where the fragment
 is formed and a coarser mesh is used far from the cloud.
We apply so-called log-mesh: the size of the grid is chosen like
 $\Delta z_i = F_z \Delta z_{i-1}$ for $i > i_{c}$,
 $\Delta z_i = F_z \Delta z_{i+1}$ for $i < i_{c}$,
 and $\Delta r_j = F_r \Delta r_{j-1}$ ($F_z$, $F_r > 1$),
 where $i$ and $j$ represent the sequential numbers of the cell in the
$z$-direction
 and in the $r$-direction, respectively, and subscript c means the
 the midpoint of the $z$-axis.
Models CB and CB2 have the same physical parameters as model B
 but they have $\simeq$ 2 - 4 times closer spatial resolution than model B,
 respectively.
Although these two models use different initial amplitude of perturbations
 (table 1),
 qualitative evolutions are similar with each other.

Due to a finer resolution,
 the evolution can be traced till the maximum density reaches
 $\simeq 10^6=10^4\times$ initial central density.
In Figure 8, we show the structure at $t=0.705$ for model CB2
 (due to a larger amplitude of $\delta_{\rm ini}=10^{-1}$ in model CB2,
  irregularities grow faster than models CB and B.
  This time scale is equivalent to $t=1.26$ for models in which
 smaller initial perturbation is assumed as $\delta_{\rm ini}=10^{-2}$).
As shown in Figure 8c and d, collapse speed becomes faster as reaching the
 fragment center.
And the maximum falling speed increases as collapse proceeds.
{}From Figure 8c, at that time, the maximum infall velocity
 in the $z$-direction agrees with that observed
 in the asymptotic solution for the spherical
 isothermal collapse by Larson(1969) and Penston(1969), $3.28 c_s$.
It is shown that high-density region with a thickness of $\sim 10^{-2}$, i.e.,
 $|z-z_c| \alt 10^{-2}H= 10^{-1} c_s/(4\pi G \rho_{c~\rm init})^{1/2}$,
 is formed, in which contraction speed is decreased as reaching $z_c$.
In the $r$-direction, in contrast, very smooth distributions of density,
 magnetic field strength, and velocity are seen.
This numerical result qualitatively agrees with the calculation
 by Hanawa, Nakamura, \& Nakano (private communication) using a different
 numerical scheme.

Thin dense object which is contracting slowly
 now seems to be separated from the outer inflow region seen in
 $|z-z_c| \agt 10^{-2}$.
The fraction of mass which form this relatively static high-density disk
 is estimated as 6\% - 10 \%.
So, if stars are formed from the relatively static matter,
 the ratio of the mass of newly formed protostars to that of a parent cloud
 becomes as 6 \% - 10 \%.

Therefore, the next evolution stage after this simulation ends
 seems to be a proto-stellar system
  consisting of a proto-star and an accreting matter onto
 the proto-star, whose luminosity comes from the liberated
 gravitational energy of falling material.
%
%
\section{Discussion}
\setcounter{equation}{0}

\subsection{A Model with Uniform Magnetic Fields}

In the preceding section, we investigated the models that the cloud
 is supported by magnetic fields.
In contrast, here, we study the model with uniform magnetic field
 and compare the structure of fragments.
Instead of $B_z^2/8\pi \propto c_s^2\rho$,
 we assume that the strength of $B_z$ is constant.
Radial density distribution is given by equation (\ref{eqn:nomag}).
Relative strength of the magnetic fields is
 expressed by a parameter $\alpha_c$ which is equivalent to the
 ratio of $B_z^2/4\pi$ to $c_s^2\rho$ at the center of the cloud.

In Figure 9, the structure at $t=1.74$ is shown.
This figure shows several differences with Figures 1 and 2
 (constant $\alpha$ model).
(i) the cloud outer boundary: the distance from the $z$-axis increases
 near $z\simeq l_z/2$, and decreases far from the fragment.
This seems to come from the fact that the magnetic fields becomes more
 important far from the $z$-axis and the gas moves toward $z\simeq l_z/2$
 without any effects on the magnetic fields (Fig.9b).
As a result, the density contours indicates a convex shape.
Contrarily, in models with constant $\alpha$, the gas falls down along the
 valley of magnetic fields.
Thus, the cylindrical tube of the cloud shrinks near $z\simeq l_z/2$
 and inflates near $z\simeq 0$ and $z\simeq l_z$.
(ii) the magnetic field line: as seen in Figure 9b
 the field line breaks weaker than previous models.
This seems to come from the fact that the configuration is not unstable
 against the Parker instability, that is, the magnetic valley does not
 necessarily promote the instability in this case.
That is, there is no positive feedback.
The second factor is: since the ratio of magnetic pressure to
 the thermal one, $\alpha/2$, increases outwardly,
 the magnetic field becomes relatively important for the dynamics
 with increasing $r$, even if $\alpha_c$ is taken $\simeq 1$.
Thus the field line is almost locked in the initial configuration.

We studied two models for the magnetic field configuration,
 $\alpha=$constant and $B_z=$constant, which correspond two extremes
 which are realized in the real interstellar space.

\subsection{Classification of the Cylindrical Cloud}

How does a cylindrical cloud with a finite line density $\lambda$ and
 a magnetic field strength evolve?
In the preceding section, it is shown that
 even in a cloud in the hydrostatic equilibrium,
 fragments are formed and dense part of the fragment falls into a runaway
 collapse.
{}From equation (\ref{eqn:fB}), the line density of a cloud
 with $F=\rho_c/\rho_s$,
 is written as
\begin{equation}
 \lambda(F)=\frac{2c_s^2}{G}
                 \frac{F^{1/2}-1}{F^{1/2}}\left(1+\frac{\alpha}{2}\right).
\end{equation}
This leads to the maximum line density which can be supported by
 the thermal and magnetic pressure as
\begin{equation}
 \lambda_{\rm max}=\frac{2c_s^2}{G}\left(1+\frac{\alpha}{2}\right),
 \mbox{\hspace{1cm}for\hspace{3mm}} F \rightarrow \infty,
\end{equation}
which include the critical line density: equation(\ref{eqn:mcr-cyl}).

Therefore, if the cloud has a line density smaller than $\lambda_{\rm max}$,
 which is related to the isothermal sound speed, $c_s$, and the Alfv\'{e}n
 speed, $\alpha^{1/2} c_s$,
 the cloud has a magnetohydrostatic equilibrium.
In such a cloud, irregularity grows in several $\times \tau_{\rm max}$
 depending on the amplitude of irregularities.
Adopting $3\tau_{\rm max}$ as a typical growth time, this gives
\begin{equation}
 \tau_{\rm grow}\simeq 2.4\times 10^6 {\rm yr}
  \left(\frac{F}{100}\right)^{-1/2}
  \left(\frac{\rho_s}{2\times 10^{-22}{\rm g~cm^{-3}}}\right)^{-1/2}.
\end{equation}
Final phase of the contraction, that is, the phase when the maximum density
 in the high-density core increases rapidly, continues only in a short duration
 $\sim 0.1 \times \tau_{\rm grow}$.

On the other hand, a cloud with super-critical line density $\lambda >
 \lambda_{\rm max}$ does not have any magnetohydrostatic configuration.
Thus, the cloud contracts as a whole, and forms a thin cloud like a string.
Recently, hydrodynamical simulation has been done for such a massive
 contracting cylindrical clouds
 by Inutsuka and Miyama (private communication) for non-magnetic clouds.
They found that in such clouds,
 the irregularity grows relatively slowly compared with
 the contraction time-scale $\simeq$ free-fall time-scale and
 the cloud does not fragment but forms a string as long as the gas
 obeys the isothermal equation of state.

Comparing these two, it is expected that
 super- and subcritical cylindrical cloud
 show completely different evolutions.
This may be related to the mass function of the new-born stars.
%
%
\section{Summary}
\setcounter{equation}{0}

We studied the process of fragmentation in an isothermal
 cylindrical cloud with infinite length.
By a numerical magnetohydrodynamical method,
 the evolution of the fragment
 was investigated from the linear stage to nonlinear stage throughout.
It is shown that the fastest growing perturbation has the same wavelength
 as predicted with the linear theory.
In a linear stage the fragment appears as a prolate spheroidal shape.
However, in a nonlinear stage, the fragment threaded by the magnetic field
 forms a disk perpendicular to the field line.
At the center of the fragment there forms a high-density core, which
 continues to collapse.
Non-magnetized cloud only forms  collapsing spherical cores which are
 separated by $\lambda_{\rm max}$.
For magnetic clouds, it is shown that
 when the collapsing velocity reaches $\sim 3.5 c_s$,
 relatively slowly contracting dense inner part is formed in the contracting
 disk,
 which also extends perpendicular to the magnetic fields.
This disk is in almost static and is separated from the accretion flow
 which is accelerated as reaching the center of the core.

\vspace{0.5cm}
I would like to thank T.~Hanawa (Nagoya University) for stimulating
 discussion.
We compared the numerical results with different schemes and could confirm
 the reliability of the schemes and the physics.
I also thank S.~Inutsuka, S.~M.~Miyama (National Astronomical Observatory) and
 M.~Y.~Fujimoto (Niigata University) for useful discussions.
Numerical calculations were carried out by supercomputers:
 {\sc Hitac} S-820/80's (Hokkaido University and University of Tokyo),
 {\sc Facom} VP200 (Institute of Space and Aeronautical Sciences).
This work was supported in part by Grants-in-Aid for Science Research
 from the Ministry of Education, Science, and Culture (04233211
 and 05217208).

\newpage
\noindent{\large {\bf Tables}}
\begin{table}[h]
\begin{center}
TABLE 1: Model Parameters

\begin{tabular}{cccccc}
\multicolumn{6}{c}{ } \\
\hline \hline
Model \ \ \ & $\alpha$ &  $F$ & $\delta_{\rm ini}$  & $l_z$ & cell number    \\
\hline
A \dotfill\ & 1        & 100  & 0         & 1.935  & 400$\times$400 \\
A2\dotfill\ & 1        & 100  & 0         & 3.87   & 400$\times$400 \\
B \dotfill\ & 1        & 100  & $10^{-2}$ & 1.935  & 400$\times$400 \\
C \dotfill\ & 0        & 100  & $10^{-2}$ & 2.21   & 400$\times$400 \\
DB$^a$ \dotfill\ & 1        & 10   & $10^{-2}$ & 1.935  & 400$\times$400 \\
DC$^a$ \dotfill\ & 0        & 10   & $10^{-2}$ & 2.21   & 400$\times$400 \\
DD$^a$ \dotfill\ & 4        & 10   & $10^{-2}$ & 1.935 & 400$\times$400 \\
CB$^b$ \dotfill\ & 1       & 100  & $10^{-2}$ & 1.935  & 400$\times$280 \\
CB2$^b$ \dotfill\ & 1       & 100  & $10^{-1}$ & 1.935  & 800$\times$560 \\
UB$^c$ \dotfill\ & 1       & 100  & $10^{-2}$ & 1.935  & 200$\times$200 \\
\hline
\end{tabular}
\end{center}

\noindent
$^a$In a series of models with character D, diffuse clouds are
 considered.

\noindent
$^b$Spatially varying spacing grid is used.
The physical parameters are taken as identical as model B.

\noindent
$^c$Uniform magnetic fields are assumed.
In this model, $\alpha$ represents the value at the center of the cloud
 ($r=0$), that is $\alpha_c$.
\end{table}

\vspace{1cm}
\begin{table}[h]
\begin{center}
TABLE A1: Necessary CPU time to solve the Poisson equation once

\begin{tabular}{llccc}
\\
\hline\hline
Machine &\hspace*{6cm} & CGS (sec) & MILUCGS (sec) & $N$ \\
\hline
Sparc IPX &(Workstation) \dotfill          &  505      &  143          & 200 \\
M-682H &(General Purpose Computer)\dotfill &   43.1    &   14.1        & 200 \\
S-820/80 &(Supercomputer) \dotfill         &    4.1    &    2.4        & 400 \\
\hline
\end{tabular}
\end{center}
\end{table}

\newpage
\noindent
{\large {\bf References}}

\re
Bastien, P. 1983, \AAp, 119, 109

\re
Bastien, P., Arcoragi, J.-P., Benz, W., Bonnel, I., \&
 Martel, H. 1991, \ApJ, 378, 255

\re
Black, D. C. \& Scott, E. H. 1982, \ApJ, 263, 696

\re
Bonner, W. B. 1956, \MN, 116, 351

\re
Brio, M., \& Wu, C. C. 1988, J. Comp. Phys. 75, 400

\re
Chandrasekhar, S. 1939, An Introduction to the Study of Stellar Structure
 (Chicago: University of Chicago Pr.), $\S$22

\re
Dongarra, J. J., Duff, I. S., Sorensen, D. C., \& van der Vorst, H. A. 1991,
 { Solving Linear System on Vector and Shared Memory Computers},
 (Philadelphia: SIAM)

\re
Dorfi, E. 1982, \AAp, 114, 151

\re
---------------. 1989, \AAp, 225, 507

\re
Ebert, R. 1955, Zs. Ap. 37, 217

\re
Elmegreen, B. G., \& Elmegreen, D. M. 1978, \ApJ, 220, 1051

\re
Evans, C. R. \& Hawley, J. F. 1988, \ApJ, 332, 659

\re
Gustafsson, I. 1978,  BIT, 18, 142

\re
Larson, R. B. 1969, \MN, 145, 271

\re
---------------. 1991, in { Fragmentation of Molecular Clouds and
 Star Formation}, ed. E. Falgarone, F. Boulanger, \& G. Duvert
 (Dordrecht: Kluwer), p.261

\re
Meijerink, J. A. \& van der Vorst, H. A. 1977,  Math. Comp., 31, 148

\re
Mouschovias, T. Ch. 1977, \ApJ, 211, 147

\re
---------------. 1979, \ApJ, 228, 159

\re
Nagasawa, M. 1987, Prog. Theor. Phys. 77, 635

\re
Nakamura, F., Hanawa, T, \& Nakano, T. 1993, \PASJ\ in press

\re
Nakano, T. 1979, \PASJ, 31, 697

\re
---------------. 1988 in Galactic and Extragalactic Star Formation,
 ed. R. E. Pudritz \& M. Fich, (Dordrecht: Kluwer), p.111

\re
Norman, M. L. \& Winkler, K.-H. A. 1986, in Astrophysical Radiation
 Hydrodynamics, ed. K.-H. A. Winkler \& M. L. Norman (Dordrecht: Reidel), p.187

\re
Parker, E. N. 1979, Cosmical Magnetic Fields (Oxford: Oxford University Pr.),
$\S$13 \&
 $\S$22

\re
Penston, M. V. 1969, \MN, 144, 457

\re
Phillips, G. L. 1986a, \MN, 221, 571

\re
---------------. 1986b, \MN, 222, 111

\re
Scott, E. H. \& Black, D. C. 1980, \ApJ, 239, 166

\re
Shu, F. H., Adams, F. C., \& Lizano, S. 1987, ARAAp, 25, 23

\re
Spitzer, L., Jr. 1978, Physical Processes in the Interstellar Medium,
 (New York: Wiley), $\S$ 11.3 \& $\S$ 13.3

\re
Tomisaka, K. 1992, \PASJ, 44, 177

\re
Tomisaka, K., Ikeuchi, S., \& Nakamura, T. 1988, \ApJ, 335, 239

\re
---------------. 1989, \ApJ., 341, 220 (errata 346, 1061)

\re
van Leer, B. 1977, J. Comp. Phys., 23, 276

\re
Weaver, R., McCray, R., Castor, J., Shapiro, P., \& Moore, R. 1977,
 \ApJ, 218, 377 (errata 220, 742)

\newpage
\appendix
\section{MILUCGS}
\setcounter{equation}{0}
Here, a brief description of MILUCGS and its performance are described.
The finite difference scheme for the Poisson equation is written
\begin{equation}
 \left(
   \begin{array}{cccccccc}
     a_1     & ub_1   &        &        & uc_1   &        &        &       \\
     db_2    & a_2    & ub_2   &        &        & uc_2   &        &       \\
             & \ddots & \ddots & \ddots &        &        & \ddots &       \\
             &        & \ddots & \ddots & \ddots &        &        & \ddots\\
     dc_{m+1}&        &        &db_{m+1}& a_{m+1}&ub_{m+1}&        &       \\
            & dc_{m+2}&       &        & \ddots & \ddots & \ddots &       \\
            &        & \ddots &        &        & \ddots & \ddots & \ddots\\
            &        &        &\ddots  &        &        & \ddots & \ddots\\
   \end{array}
 \right)
 \left(
   \begin{array}{c}
     \psi_1\\
     \psi_2\\
     \psi_3\\
     \vdots\\
     \vdots\\
     \vdots\\
     \vdots\\
     \psi_{nm}
  \end{array}
 \right)
   =
 \left(
   \begin{array}{c}
     \rho_1\\
     \rho_2\\
     \rho_3\\
     \vdots\\
     \vdots\\
     \vdots\\
     \vdots\\
     \rho_{nm}
  \end{array}
 \right),
 \label{A1}
\end{equation}
where two-dimensional expression of $\psi_{i,j}$ and $\rho_{i,j}$
($1\le i\le m$ and $1\le j\le n$) are converted to a one-dimensional
expression $\psi_k$ and $\rho_k$ with $k=(j-1)\times m+i$.
$m$ and $n$ represent, respectively, the cell numbers in the $z$- and
$r$-direction.
The coefficients $db$ and $ub$ correspond to the derivatives with
 respect to the $z$-direction, and the coefficients of $dc$ and $uc$
 correspond to those of $r$-direction.
If the problem is a simple Neuman or Dirichret boundary condition,
 the matrix to be solved is like equation (\ref{A1}).
However, since we apply here a cyclic boundary in the $z$-direction,
 we have to add two coefficients expressing this boundary condition.
That is, the term of $\psi_{m,j}=\psi_{jm}$ should connect with
 $\psi_{m-1,j}=\psi_{jm-1}$ and $\psi_{1,j}=\psi_{jm-m+1}$ instead
 of $\psi_{jm+1}$.
The equation for cyclic boundary becomes
\begin{equation}
 \left(
   \begin{array}{cccccccc}
     a_1     & ub_1   &        & ud_1   & uc_1   &        &        &       \\
     db_2    & a_2    & ub_2   &        & ud_2   & uc_2   &        &       \\
             & \ddots & \ddots & \ddots &        & \ddots & \ddots &       \\
     dd_{m}  &        & \ddots & \ddots & \ddots &        & \ddots & \ddots\\
     dc_{m+1}&dd_{m+1}&        &db_{m+1}& a_{m+1}&ub_{m+1}&        & \ddots\\
            & dc_{m+2}&dd_{m+2}&        & \ddots & \ddots & \ddots &       \\
            &        & \ddots &\ddots   &        & \ddots & \ddots & \ddots\\
            &        &        &\ddots   &\ddots  &        & \ddots & \ddots\\
   \end{array}
 \right)
 \left(
   \begin{array}{c}
     \psi_1\\
     \psi_2\\
     \vdots\\
     \vdots\\
     \psi_{m+1}\\
     \vdots\\
     \vdots\\
     \psi_{nm}
  \end{array}
 \right)
   =
 \left(
   \begin{array}{c}
     \rho_1\\
     \rho_2\\
     \vdots\\
     \vdots\\
     \rho_{m+1}\\
     \vdots\\
     \vdots\\
     \rho_{nm}
  \end{array}
 \right),
 \label{A2}
\end{equation}
where only $dd_{j\times m}$ as well as $ud_{(j-1)m+1}$ has non-zero
 components and the others have the value of zero.
This linear simultaneous equation is solved by conjugate gradient squared
method (Dongarra \etal 1991).
If we write equation(\ref{A2}) as $A\hat{\psi}=\rho$,
 the algorithm is as follows:
\begin{enumerate}
 \item Prepare the initial guess $\hat{\psi}_0$\\
	$\hat{r}_0=\rho-A\hat{\psi}_0$\\
	$\hat{p}_0=e_0=\hat{r}_0$\\
	$l=0$
 \item while $\| \hat{r_l}\| > \epsilon \|\rho\|$ do
 \begin{enumerate}
  \item $\alpha_l=(\hat{r}_0,\hat{r}_l)/(\hat{r}_0,A\hat{p}_l)$
  \item $h_{l+1}=e_{l}-\alpha_lA\hat{p}_l$
  \item $\hat{r}_{l+1}=\hat{r}_l-\alpha_lA(e_l+h_{l+1})$
  \item $\hat{\psi}_{l+1}=\hat{x}_l+\alpha_l(e_l+h_{l+1})$
  \item $\beta_l=(\hat{r}_0, \hat{r}_{l+1})/(\hat{r}_0,\hat{r}_l)$
  \item $e_{l+1}=\hat{r}_{l+1}+\beta_lh_{l+1}$
  \item $\hat{p}_{l+1}=e_{l+1}+\beta(h_{l+1}+\beta_l\hat{p}_l)$
  \item $l=l+1$.
 \end{enumerate}
\end{enumerate}
The convergence of this method is improved much by preconditioning.
In the present paper, we apply the incomplete LU-decomposition.
The incomplete LU-decomposition is a method that matrix $A$ is
decomposed as $LDU-R$, where $L$, $D$, $U$, and $R$ are a lower triangle
matrix, a diagonal matrix, an upper triangle matrix, and a residual matrix.
Instead of $A\hat{\psi}=\rho$, we apply the CGS method to
the reduced equation of
\begin{equation}
(LDU)^{-1}A\hat{\psi}=(LDU)^{-1}\rho.
\label{A3}
\end{equation}
Since $(LDU)^{-1}A$ is nearly equal to the identity matrix $I$
 and the eigen values of this matrix gather around unity,
 the convergence of the CGS method is much accelerated.
The way to decompose $A=LDU-R$ is arbitrary.
Here, we use the method by Meijerink \& van der Vorst (1977) and
Gustafsson (1978).
Describing briefly, $L$, $D$, and $U$ are chosen as
\begin{equation}
 L=
 \left(
   \begin{array}{cccccccc}
1/\tilde{d}_1&        &        &        &        &        &        &      \\
     db_2    &1/\tilde{d}_2&   &        &        &        &\bigzerou&      \\
     \vdots  & \ddots & \ddots &        &        &        &        &       \\
     dd_{m}  &        & \ddots & \ddots &        &        &        &       \\
     dc_{m+1}&dd_{m+1}&      &db_{m+1}& 1/\tilde{d}_{m+1}&&        &       \\
            & dc_{m+2}&dd_{m+2}&        & \ddots & \ddots &        &       \\
            &        & \ddots &\ddots   &        & \ddots & \ddots &       \\
            &        &        &\ddots   &\ddots  &&db_{mn}&1/\tilde{d}_{mn}\\
   \end{array}
 \right),
\end{equation}
\begin{equation}
D=
 \left(
   \begin{array}{cccccccc}
  \tilde{d}_1&        &        &        &        &        &        &      \\
        &\tilde{d}_2  &        &        &        &        &\bigzerou&     \\
             &        & \ddots &        &        &        &        &      \\
             &        &        &  \ddots&        &        &        &      \\
             &        &        &        & \ddots &        &        &      \\
             &        &        &        &        & \ddots &        &      \\
            &\bigzerol&        &        &        &    &\tilde{d}_{mn-1}&  \\
             &        &        &        &        &        &&\tilde{d}_{mn}\\
   \end{array}
 \right),
\end{equation}
\begin{equation}
 U=
 \left(
   \begin{array}{cccccccc}
1/\tilde{d}_1& ub_1   &        & ud_1   & uc_1   &        &        &      \\
       & 1/\tilde{d}_2& ub_2   &        & ud_2   & uc_2   &        &      \\
             &        & \ddots & \ddots &        & \ddots & \ddots &       \\
             &        &        & \ddots & \ddots &        & \ddots & \ddots\\
             &      &    &    & 1/\tilde{d}_{m+1}&ub_{m+1}&        & \ddots\\
             &        &        &        &        & \ddots & \ddots &       \\
            &\bigzerol&        &        &        &        & \ddots & \ddots\\
             &        &        &        &        &        &        & \ddots\\
   \end{array}
 \right).
\end{equation}
If we choose the diagonal element of $D$ as
\begin{equation}
  \tilde{d}^{-1}=a_k-db_k \tilde{d}_{k-1} ub_{k-1}
     - dc_k \tilde{d}_{k-m} uc_{k-m}
     - dd_k \tilde{d}_{k-m+1} ud_{k-m+1},
\end{equation}
the diagonal part of the matrix $A$ becomes equal to that of $LDU$.
However, when the matrix $LDU$ is operated on $\hat{\psi}$, it
 generates extra terms (fill-in) such as
\begin{equation}
  db_k \tilde{d}_{k-1} uc_{k-1} \hat{\psi}_{k+m-1}+
  dc_k \tilde{d}_{k-m} ub_{k-m} \hat{\psi}_{k-m+1}
\end{equation}
as well as the ordinary terms:
\begin{equation}
  a_k \hat{\psi}_{k} + ub_k \hat{\psi}_{k+1} + db_k \hat{\psi}_{k-1}
 + uc_k \hat{\psi}_{k+m} + dc_k \hat{\psi}_{k-m}
 + ud_k \hat{\psi}_{k+m-1} + dd_k \hat{\psi}_{k-m+1}.
\end{equation}
To suppress the effect of these extra terms, the diagonal term $\tilde{d}$
is modified as
\begin{equation}
  \tilde{d}^{-1}=a_k-db_k \tilde{d}_{k-1} ub_{k-1}
               - dc_k \tilde{d}_{k-m} uc_{k-m}
               - dd_k \tilde{d}_{k-m+1} ud_{k-m+1}
                - \gamma(db_k \tilde{d}_{k-1} uc_{k-1}
                        +dc_k \tilde{d}_{k-m} ub_{k-m}),
\end{equation}
where the factor $\gamma$ is chosen as the way by Gustafsson (1978).
Using the LDU-decomposition, in the CGS algorithm the product
 of $A v$ is replaced with $(LDU)^{-1}Av$.
The operation multiplying $(LDU)^{-1}$ is done as follows:
 $y=(LDU)^{-1}x$ is equivalent to $x=(LDU)y$, i.e.,
\begin{equation}
 Lz=x
 \label{Lz}
\end{equation}
\begin{equation}
 DUy=z.
 \label{DUy}
\end{equation}
The equation(\ref{Lz}) is solved as: for $k=1,2,\cdots,mn$
\begin{equation}
 z_k=\tilde{d}_k(x_k-db_k z_{k-1}- dc_k x_{k-m} -dd_k x_{k-m+1}),
\end{equation}
and then equation (\ref{DUy}) is solved as: for $k=mn,mn-1,\cdots,1$
\begin{equation}
 y_k=z_k-\tilde{d}_k(ub_k y_{k+1}+ uc_k y_{k+m} +ud_k y_{k+m-1}).
\end{equation}

Necessary CPU time is proportional to the cell number, $mn$, and
 to loop count.
A numerical experiment was done and we compared CGS with MILUCGS.
The adopted test problem is as follows:
 the gravitational potential for a uniform cylinder is solved
 with square grids of $m=n=N$.
The same problem was solved by a supercomputer Hitachi S-820/80,
 a general purpose main frame machine Hitachi M-682H, and a
 workstation Sun Sparc IPX.
The allowable maximum relative error is chosen as $\epsilon=10^{-7}$
 and calculations are done using 64 bits.
As a result, about $80\times(N/200)$ cycles are necessary to achieve a solution
 for MILUCGS,
 while the simple CGS requires $410\times(N/200)$.
Since the number of cycles necessary for convergence is proportional
 to $N$, CPU time is proportional to $N^3$ irrespective with
 the preconditioning or not.

In table A1, we summarize the CPU time necessary for solving the
 Poisson equation by respective machines.
Comparing CPU time, the preconditioning reduces the CPU time about
 $\alt 1/3$
 for scalar machines (M-682 and Sparc IPX).
Since the amount of operations in MILUCGS is larger than that of CGS,
 the factor is not equal to $\simeq1/5$ (the ratio of necessary loops)
 but to $\alt 1/3$.
However, this clearly shows the advantage of MILUCGS over the simple CGS.
The difference for the supercomputer is relatively small,
 for in order to vectorize the operation of multiplying $(LDU)^{-1}$
 [equations (\ref{Lz}, \ref{DUy})]
 a more complicated algorithm with list vectors is necessary.

\newpage
\noindent
{\large {\bf Figure Captions}}

\re
{\bf Fig.1}: The initial radial density distributions.
A solid line corresponds to models A and B, i.e., a magnetized cloud with
 $\alpha=1$.
A dashed line shows the density of model C which has no magnetic fields
 $\alpha=0$.
Both  correspond to the models with the center-to-surface density ratio,
 $F=100$.
The dotted part of the curves represents the ambient medium $f < 1$ which
 is assumed to have no effect as a source of gravity.

\re
{\bf Fig.2}: The time evolutions of density (left panel) and
 magnetic fields (right panel) for model A.
The $z$-axis and $r$-axis are directed horizontally and vertically,
 respectively, in contrast to an ordinary fashion.
The number attached to the density contour lines represent
 the logarithm of the density: $\log_{10} \rho$.
The step of the contour lines is taken constant $=0.25$.
Four snapshots, i.e., $t=2.4$ (a), $t=3.6$ (b),
 $t=4.9$ (c), and $t=5.6$ (d) are shown.
In $t=3.6$, no prominent fragmentations appear.
In the stage shown in (c), a high-density region elongating
 in the $z$-direction appears.
After (c), contraction along the symmetry axis proceeds.
The final structure is a disk contracting towards the center seen in (d).
The disk is perpendicular to the symmetric axis of the cylinder.

\re
{\bf Fig.3}: The structure of the cloud for model B.
The upper-left panel shows the density distribution (a) and
 the upper-right does the magnetic field lines (b).
It is shown that a fragment contracts and forms a disk perpendicular to the
 symmetric axis.
Cross-cut views along the $z$-axis (c) and the $r$-axis (d) are plotted.
In panel c, a logarithmic plot of $\rho(z,0)$ and $B_z(z,0)$ and
 a linear plot of $v_z(z,0)/10$ are illustrated.
Similarly, in panel d, a logarithmic plot of $\rho(z_c,r)$ and $B_z(z_c,r)$ and
 a linear plot of $v_r(z_c,r)/10$ are shown,
 where $z_c$ represents the position of the center of the fragment $=l_z/2$.
This snapshot corresponds to the state of $t=1.26$.

\re
{\bf Fig.4}: Time evolutions of the relative density enhancement,
 $\delta_{\rm max}$, for models A and B.
This is defined as
$
 \delta_{\rm max} \equiv \rho(r=0)|_{\rm max}/<\rho(r=0)>-1
$.
The curve for model B is similar to the late phase ($t \agt 4$)
 of the model A.
This shows that the structure and evolution of the fragment
 can be calculated from the initial state which has the
 most unstable perturbation with a finite amplitude.

\re
{\bf Fig.5}: The time evolution of density for Model C.
In this model, the fragmentation in non-magnetic cloud is
 studied.
Since the magnetic field, which has the effect of supporting
 cloud laterally, is not included, the radial extent of the
 cloud is thinner than that of models A and B.
Two snapshots of density distribution
 at $t=1.2$ (a) and $t=1.4$ (b) are plotted.

\re
{\bf Fig.6}: The cross-cut views of the cloud of models B (a) and C (b).
The left panel shows the variation of $\rho(z,0)$
 against $\log|z-z_c|$, and the right panel
 shows that of $\rho(z_c,r)$ against $\log r$.
Two characteristic solutions [(\ref{eqn:singularS'}), and
 (\ref{eqn:singularC'})] are also shown in right panels.
Five snapshots at $t=0.3$, 0.6, 0.9, 1.2, 1.4 are plotted.

\re
{\bf Fig.7}: The density distributions of models DB (a), DD (b), and DC (c).
Model DB ($\alpha=1$) has the same parameter as model B except for the density
 contrast between the center and the surface, $F=10$.
Model DD corresponds to the cloud with stronger magnetic fields ($\alpha=4$).
Model DC ($\alpha=0$) has the same parameter as model C except for the density
 contrast between the center and the surface, $F=10$.
For panel c, comparing how the density contour lines are running with
 that for Fig.5, it can be seen that the density distributions
 are very similar with each other.
Since the length of numerical box in the $z$-direction $l_z$ is
 taken as proportional to $\rho_c^{-1/2}$,
 it is seen that the fragments of models C and F have an almost
 similar structure, if the size is scaled with the ``scale-height''
 at the center, $c_s/(4\pi G \rho_c)^{1/2}$ instead of $H$.
The number attached to the contour line, $k$, indicates that the value
 of the contour is equal to $\rho=10^{k/4}$.

\re
{\bf Fig.8}: The structure of the fragment of model CB2 at $t=0.705$.
Physical parameters of this model are chosen identical with model B but
 this model has 4.28 times higher spatial resolution than model B.
In this model a perturbation with a larger amplitude $\delta_{\rm ini}=10^{-1}$
 is assumed initially to reduce a computing time.
This time corresponds to $1.26$ for model CB with $\delta_{\rm ini}=10^{-2}$.
The panel a shows the density and velocity fields and
 b shows the magnetic field lines.
To see the structure near the fragment clearly,
 in panel a, only a region of $0<z<l_z/2$ and $r<l_z/2$ is plotted.
Velocity vectors are plotted every 16 grids.
Panels c and d are respectively, the cross-cut views along the $z$-axis
 and $r$-axis.
A logarithmic plot of $\rho(z,0)$ and $B_z(z,0)$ and
 a linear plot of $v_z(z,0)$ are illustrated against $\log |z-z_c|$ in panel c.
Similarly, in panel d,
 a logarithmic plot of $\rho(z_c,r)$ and $B_z(z_c,r)$ and
 a linear plot of $v_r(z_c,r)$ are shown against $\log r$.

\re
{\bf Fig.9}: The structure of the fragment formed in a cylindrical
 cloud threaded by a uniform magnetic field (model UB).
The initial radial distribution of the density is as the same as model C
 (non-magnetic model), since the magnetic fields do not play a role in
 supporting the cloud.
The density contour (a) and the magnetic fields (b) are illustrated.
Panels c and d are, respectively, the cross-cut views along the $z$-axis
 and the $r$-axis.
That is, a logarithmic plot of $\rho(z,0)$ and $B_z(z,0)$ and
 a linear plot of $v_z(z,0)$ are illustrated in panel c.
Similarly, in panel d, a logarithmic plot of $\rho(z_c,r)$ and $B_z(z_c,r)$ and
 a linear plot of $v_r(z_c,r)$ are shown.
This snapshot corresponds to the state of $t=1.26$.

\end{document}